\documentclass[sigconf]{acmart}

\usepackage{graphicx}

\usepackage{acronym}

\usepackage{mdframed}
\usepackage{framed}

\usepackage{balance}
\usepackage{hyperref}
\usepackage{mathptmx}
\usepackage{amsmath}
\usepackage{color,soul}

\copyrightyear{2022} 
\acmYear{2022} 
\setcopyright{acmlicensed}
\acmConference[NSPW 2022]{FULL CONF NAME}{DATE}{LOCATION}
\acmBooktitle{}
\acmPrice{15.00}
\acmDOI{}
\acmISBN{}

\graphicspath{ {./images/} }

\newenvironment{finding}{\begin{framed}\noindent\textbf{Gaps.}~}{\end{framed}}
\newenvironment{research}{\begin{framed}\noindent\textbf{Agenda.}~}{\end{framed}}
\newenvironment{moral}{\begin{framed}\textbf{Moral.}\noindent}{\end{framed}}
\newenvironment{significance}{\begin{framed}\noindent\textbf{Significance.}~}{\end{framed}}
\acrodef{SoK}{Systematization of Knowledge}
\acrodef{HCI}{Human Computer Interaction}
\acrodef{DCS}{Developer-Centered Security}
\acrodef{PETs}{Privacy Enhancing Technologies}
\newcommand{\etal}[0]{et~al{.}}

\begin{document}

\title{From Utility to Capability: A New Paradigm to Conceptualize and Develop Inclusive PETs}
% \author{{Partha Das Chowdhury, Andrés Domínguez Hernández, Kopo M. Ramokapane, Awais Rashid}\\{Bristol Cyber Security Group}\\ 
% {University of Bristol, UK}\\{\{firstname.lastname\}@bristol.ac.uk}}

\author{Partha Das Chowdhury}
\email{partha.daschowdhury@bristol.ac.uk}
\affiliation{%
    \department{Department of Computer Science}
    \institution{University of Bristol}
    \country{United Kingdom}
}    
    \author{Andrés Domínguez Hernández}
\email{andres.dominguez@bristol.ac.uk}
\affiliation{%
    \department{Department of Computer Science}
    \institution{University of Bristol}
    \country{United Kingdom}
}
\author{Kopo M. Ramokapane}
\email{marvin.ramokapane@bristol.ac.uk}
\affiliation{%
    \department{Department of Computer Science}
    \institution{University of Bristol}
    \country{United Kingdom}
}

\author{Awais Rashid}
\email{awais.rashid@bristol.ac.uk}
\affiliation{%
    \department{Department of Computer Science}
    \institution{University of Bristol}
    \country{United Kingdom}
}

\date{}

\begin{abstract}
The wider adoption of \ac{PETs} has relied on usability studies – which focus mainly on an assessment of how a specified group of users interface, in particular contexts, with the technical properties of a system. While human-centred efforts in usability aim to achieve important technical improvements and drive technology adoption, a focus on the usability of \ac{PETs} alone is not enough. \ac{PETs} development and adoption requires a broadening of focus to adequately capture the specific needs of individuals, particularly of vulnerable individuals and/or individuals in marginalized populations. We argue for a departure, from the utilitarian evaluation of surface features aimed at maximizing adoption, towards a bottom-up evaluation of what real opportunities humans have to use a particular system. We delineate a new paradigm for the way \ac{PETs} are conceived and developed. To that end, we propose that Amartya Sen’s \emph{capability approach} offers a foundation for the comprehensive evaluation of the opportunities individuals have based on their personal and environmental circumstances which can, in turn, inform the evolution of \ac{PETs}. This includes considerations of vulnerability, age, education, physical and mental ability, language barriers, gender, access to technology, freedom from oppression among many important contextual factors.

\end{abstract}

\maketitle

\section{Introduction}\label{intro}
While privacy has been recognised as a fundamental right, there has been debate as to whether technical and regulatory interventions adequately allow everyone, irrespective of their circumstances, to exercise this right~\cite{guberek2018}. This is tied to the fundamental question of how privacy protection mechanisms are conceived, or the assumptions considered, when designing and building such mechanisms. For instance, \ac{PETs} have often seen the \emph{human} at the other end of systems as some passive \emph{user}~\cite{vemou_classification_2013}. This notion (of \ac{PETs}) does not usually account for people’s diverse interactions (or lack of) with technology and their vulnerabilities, short or long-term social, political, and economic circumstances. Neither does it grapple with the multiple ways people can be re-identified, profiled and harmed. However, not reflecting on human's diverse realities while developing such systems, may not only hinder adoption due to technical misfit but may be unintentionally harmful~\cite{salma2017}.

The \ac{HCI} and usable security communities have made a strong case for putting humans at the heart of systems design~\cite{schlesinger2017intersectional,adamss99,dodier2017paternalistic} through cross-pollination with the social sciences and experimentation with participatory and co-design methods. Other fields, for example, security economics, have also made considerable progress in widening the discussions surrounding the end users~\cite{odylzko2003,acquisiti2016,acquisitilikes2021}. While these efforts are recognized in other areas of technological development, the research community behind \ac{PETs} has remained mostly concerned with corrective mechanisms and questions of usability~\cite{ruoti2019, vemou_classification_2013, coopamootoo}. For example, Coopamootoo conducted a study to understand the use of privacy methods and highlighted that \emph{ease of use} is one of the important barriers behind low adoption of advanced \ac{PETs}~\cite{coopamootoo}. But focusing on low adoption and \emph{ease of use} as the main problems, suggests the answer to better PETs is to improve the surface features of their use~\cite{ruoti2019}. When we go beyond the frame of usability, we see other diagnoses of the inadequacy of \ac{PETs} such as Vemou~\etal{}, who identify that \ac{PETs} are not adequately sensitive to diverse cultures~\cite{vemou_classification_2013}.

The pervasive digitization of society requires everyone to engage with digital services, but it also poses the risk of harm as a result of data collection and aggregation, whether or not this is transparent to users. We argue that there is a need to expand the focus of PETs in order to account for the diverse relationships with technology and realities of age, education, dis/abilities, vulnerability, gender, race, class, marginalized individuals, migrants, and socio-political situations~\cite{senrational1994}. This is important to avoid harm, exclusion and negative impact on participation in the digital economy by vulnerable individuals. Equitable access to protection mechanisms will enable individuals to enjoy basic protections and human rights -- particularly important for vulnerable people. For example, migrants should be able access healthcare without being exposed to data misuse and exploitation.  

We contribute to this area by not only discussing the limitations of current approaches to building \ac{PETs} but also proposing a foundation upon which designers and researchers can build \ac{PETs} for everyone. Drawing on the work of Amartya Sen~\cite{sentanner}, we posit that a \emph{capability approach}\footnote{Amartya Sen articulated the \emph{capability approach} first in Tanner lectures on Human Values, delivered at Stanford University in 1979. Available on Tanner Lectures website, reprinted in John Rawls et al., Liberty, Equality and Law (Cambridge: Cambridge University Press, 1987)} based evolution of \ac{PETs} will enable individuals to achieve privacy in a manner they are able to, and they deem valuable. The \emph{capability approach} brings an evaluative space to systematically assess individuals' opportunities to live a private life by putting their freedom of choice at the heart of that assessment. If we view \ac{PETs} as a tool or service that enables the functioning of a private life, then merely possessing the service will not enable the functioning. There are other key factors such as skill, intelligence, physical ability, as well as social, economic and political circumstances. Exercising privacy, like any social good, would be dependent on these factors and circumstances~\cite{seninequality}~\footnote{For example, transport as a social good is understood and used differently by different people. For someone without legs, a standard bicycle can never be an effective means of transport, and offering a cycle would be inadequate, to say the least~\cite{crocker_robeyns_2009}.}. 

%With respect to \ac{PETs}, the significance of the \emph{capability approach}, is 
A \emph{capability approach} asks what information is necessary to make an evaluative judgment of personal, social, economic and political circumstances for the provision of \ac{PETs} for diverse social groups and marginalized sections in particular. Such an evaluation aims to inform the building of systems that are appropriate for the particular group. Further, a capability-based provisioning of PETs benefits from capturing the diversity of stakeholders' expectations of any defined system~\cite{bruce2013,rashid2016}. This notion is a departure from the resource (only) based view, which assumes that having \ac{PETs} will allow everyone to have a private life~\cite{ruoti2019}. The specific relevance of the capability-based provisioning of PETs can be further enforced by the observed diversity in expectations from and commitments of various stakeholders of any defined system~\cite{bruce2013,rashid2016}. We outline an interdisciplinary research agenda based on our proposition and invite empirical and methodological dialogue toward conceptualizing and developing more inclusive and human-centric PETs.  

%This paper is structured as follows. In Section~\ref{forwhom}, we make a case for departing from the prevailing view of usability as the foremost criterion for systems design, and in Section~\ref{reactions} we discuss the negative consequences of developing systems without a clear understanding of the diverse realities of individuals. Then, in Section \ref{formal} we give a brief overview of the \emph{capability approach}. Based on this, we propose a research agenda and outline key themes and methodologies in Section~\ref{agenda} which is followed in Section~\ref{why}, by a case implication justification connecting \emph{capability approach} in the light of some relevant examples. This is followed by conclusions in Section~\ref{conclusion}. We duly acknowledge prior studies, that highlight the shortcomings of utilitarian views of usability, but they fall short of proposing a comprehensive approach to design inclusive systems.multiple formulations of well-being (i.e., beyond utilitarian views) sought by individuals and 

\section{Interrogating Human centricity: the need to move from Utility to Capability}\label{forwhom}

Individuals should be able to use \ac{PETs} in a manner that they \emph{can} and that they \emph{value} -- intrinsic to this \emph{can} and \emph{value} is human diversity. Human diversity is the fundamental reason we propose \emph{capability approach} as a foundation on which to build \ac{PETs} -- it adequately accounts for diverse human capabilities as well as choice. While the \emph{capability approach} would evaluate individuals in diverse personal and environmental situations to point to the opportunities they have, it will do so against the minimum basic protections that every individual should have -- for example, protection against unauthorized information disclosure. Evolution of the list of minimum basic protections as well as evaluation of diverse capabilities in enjoying those protections are integral components of the \emph{capability approach} framework. This two pronged approach is a shift from the traditional utilitarian\footnote{In economics, \emph{utility} has been viewed as preference ordering – the satisfaction derived by an individual from an increased share of a good and its evaluation~\cite{sen1991utility}.} usability evaluations of system surface features with respect to a priori defined users. 

Evaluation of capabilities will capture information on the discrimination and deprivation of vulnerable individuals. The need to capture this is being acknowledged in the literature: McDonald~\etal{} argues that standardizing and advantaged lenses can impair conceptualizations of identities and the privacy needs of vulnerable people, therefore, urges that the HCI community go beyond individuals’ perception of risk to consider other conditions (e.g., power structures) that perpetuate the privacy needs of individuals~\cite{mcdonald2020privacy}; Wong~\etal{} surveyed prominent literature to understand how systems design is considered with regard to privacy in HCI. They advance the view that in order to realize privacy beyond \emph{solving, supporting and informing}, the privacy by design community must acknowledge privacy systems as socio-technical systems~\cite{wongdesignchi}; McDonald~\etal{} describe the shortcomings in commonly used privacy theories in the HCI literature which do not capture individuals who do not adhere to commonly accepted norms or expectations about users. It is challenging to thoroughly capture information about vulnerabilities through the evaluation of surface features. Because of this, some authors argue for moving from norms compliance (c.f. Nissenbaum's contextual integrity~\cite{nissenbaum2004privacy}) to human vulnerabilities -- drawing from, and building upon, feminist, queer-Marxist, and intersectional approaches to inform system design~\cite{mcdonald2020politics}. 

There have been notable attempts to improve usability by attuning design to the plural lived experiences of users. For example, Hertzum~\cite{hertzum-usability} argues for treating usability as a `sensitizing concept' rather than a predefined concept. This approach invites system designers to consider alternative variants --`images'-- of usability such as \emph{universal usability, situational usability, perceived usability, hedonistic usability, organizational usability} and \emph{cultural usability}~\cite{hertzum2010images}.  Another example is value sensitive design~\cite{friedman1996value,hendry2021value}, which offers a framework for bringing in social and ethical considerations. Sensitizing design to lived experiences and moral values are powerful in the abstraction they offer to align systems use with situated, emotional and subjective experiences. The argument we make here is that for these powerful abstractions to be effective there needs to be methodological capturing of individuals and their circumstances. Utilitarian logic of systems design, however well-meaning, cannot completely capture adequate information required to design inclusive systems. Sen, in his seminal work on equality, questions the ability of \emph{utility} to capture the concept of "needs"~\cite{seninequality}\footnote{Moreover, in \emph{An Introduction to the Principles of Morals and Legislation}, Bentham also highlights the inadequacy of the word \emph{utility} in conveying \emph{interests} and \emph{circumstances}~\cite{bentham1970introduction}}. To be sure, this is not to say that usability is irrelevant nor that higher adoption necessarily implies more inclusive systems (as conceptualized in this paper). We propose that to be inclusive --as in not discriminatory or exclusionary-- systems design should refrain from privileging usability and give independent consideration to capabilities~\cite{cranorhumans,renaud2022accessible}.

Independent consideration of capabilities means being sensitive to the social practices of individuals in difficult situations--for example refugees with diverse social practices, as well as their relationship with technology that captures their information. Rikke Bjerg~\etal{} engaged over 89 refugees in the process of settling in a new country using a digital re-settlement process. In their paper, the authors lament the HCI community's focus on usability of platforms for provisioning information to migrants -- arguing that the centering on surface technical features `exacerbates the existing barriers in the re-settlement process'. 

The \emph{Capability approach} can be used to make interpersonal comparisons of welfare to understand the comparative practical consequences of systems on individuals. Endeavors (like usability images and value sensitive design) to make \ac{PETs} sensitive need an assessment of the distributive implications of technology. Coles-Kemp~\etal{} conducted a study with 132 \emph{newcomers} seeking re-settlement in a new country, employing Ribot and Peluso's~\cite{ribot2003theory} theory of access to examine the experiences of participants with digital communications and interactions. They report that, although the HCI community emphasized fit for use and reducing cognitive load, the realization of benefit has not been comprehensively addressed. Systems designed for people in precarious situations should foreground considerations of realization and enablement rather than protection of the system~\cite{coles-kempreralisation}. The assessment of realization of benefit is tied to distributive considerations of justice.  

%This assessment will factor in personal and environmental situations like the tendency of big corporations to weave privacy narratives in a manner that effectively favors extractivist data practices~\cite{mcdonaldsm2021}.

While we argue for moving beyond the utilitarian bias, we do not discount the presence of other factors that negatively affect the adequacy of privacy protection. While the data extraction incentives~\cite{mcdonald2020politics}, and utilitarian approaches of technology evaluation~\cite{rikkebjergcivicdigital} are prominent barriers at the supply side, there are also well-known constraints at the demand side such as privacy literacy~\cite{harborth2020privacy} and accessibility~\cite{renaud2022accessible}. The successful adoption of \ac{PETs} has been found to depend not only on factors linked with technical fitness but also on users' understanding of benefits and risks, and the access-ability of their intended users~\cite{rannenberg2020,renaud2022accessible}. For example, previous work has shown that individual perception of risks are critical factors in the adoption of privacy protection technology~\cite{bada2019cyber}, which is further exacerbated by the value individuals are willing to trade for risks they do not perceive or cannot assess~\cite{akerlof1970,anderson2001information}. 
These observed diversities in risk evaluation and awareness often lead to blanket assertions like individuals do not value privacy leading to ``victim blaming"~\cite{acquisiti2004,barnes2006privacy}. Sen emphasizes the fundamental role these observed diversities play in assessing the capability of individuals~\cite{sen1992}. Thus, recognition of diversities in risk perception, awareness and privacy literacy should also be incorporated in to the evaluation of opportunities.

In fine, there is wide recognition in the literature that the mere possession of a tool is not enough for individuals to benefit from the tool -- it depends on their health, education, circumstance and other dispositions. An adequate assessment of these dispositions determine the opportunities that individuals have -- we propose \emph{capability approach} as the framework to conduct this assessment for building effective \ac{PETs}. This will allow us to build \ac{PETs} based on opportunities.

\begin{significance}
 \noindent
 Usability is concerned with how individuals interact with surface features and information provisioning (goods). This has shortcomings since humans differ in their health, ability, education and/or can be in vulnerable situations, displaced from their homes and/or living under oppressive regimes. This diversity of circumstance can have a constraining impact on marginalized and/or vulnerable individuals with respect to their use of technology and eventually exclude them. Therefore, there is a need for more adequate approaches to capture human diversities, deprivations, preferences and design systems. 
 
 %This why we argue for a paradigm shift and situate \emph{capability approach} as a foundation of how systems are conceived, designed and developed.
\end{significance}

\section{Consequences of Discounting Individual Realities}\label{reactions}
Conventional privacy protections have not not been sufficiently sensitive to the personal, and social circumstances of people.
The prevailing view of \emph{privacy as confidentiality} focuses exclusively on protecting data which is considered personal or sensitive, with the assumption that users will have adequate skills to protect their data~\cite{gurses_pets_2010}. While we acknowledge the research towards engineering confidentiality~\cite{gurses2011engineering}, this view needs to be expanded to account for the myriad ways in which people's privacy may be compromised through other means of identification. 
%We again draw upon the emphasis on socio-technical systems by Wong~\etal in the context of engineering privacy incorporating interests and contexts~\cite{wongdesignchi}. 
One might for instance draw attention to the growing ethical concerns over the use of people's digital traces, including seemingly innocuous data, for purposes of behaviour prediction and nudging, cross-referencing, profiling and policing~\cite{van_der_sloot_regulating_2020, brayne_big_2017}. In the next subsections, we outline some of the consequences of discounting individual realities manifesting in adverse behaviours and harms linked with the use of information systems.

\subsection{Information Overload and Asymmetries}\label{asymmetry} 
Citizens are confronted on a daily basis with myriad data transactions with information systems, yet very little is known about what goes on in the background or what exactly are the quid pro quos of such transactions. How data is collected, transmitted and processed by information systems has remained largely concealed~\cite{chaum1985}. The way this has been dealt with has been primarily via consent mechanisms and the publication of complex privacy policies. Yet such measures are insufficient given that concerns with the negative consequences of data sharing with commercial and governmental entities can lead people to find workarounds or experience adverse reactions including fear, reticence and feelings of resignation about participating in online activities~\cite{draper_corporate_2019,pink_data_2018}. We elaborate some of these here.
%The technologies to cater to diverse individual realities either based on narrow conceptions of privacy or they are difficult to use by a vast majority of users~\cite{salma2017,ruoti2019,gurses_pets_2010}.

\paragraph{Resignation}
A widespread assumption among service providers is that individuals irrespective of their circumstances, will be able to process complex legal jargon~\cite{ben2019data} and take an informed decision. But shifting the burden of obtaining informed consent to citizens is at odds with adequately empowering them to take control~\cite{anthonysamy2013social,fabianel2017}. In fact, this has led to a regime of misinformed and often coercive consent~\cite{millett2001}. Clear evidence of this is in the proliferation of purposefully misleading consent controls or \emph{dark patterns}~\cite{acquisitilikes2021, luguri_shining_2021}. Such asymmetrical relations engender different adverse behaviours and feelings of resignation among users. Previous research has shown that there are disconnects between the stated privacy policies and the controls that are used to implement them~\cite{anthonysamy2011}. Even if a trained individual is able to navigate through complex policies, there might not be adequate controls to enable the functioning of a private life. Service providers take advantage of people's need to access online services and their inability to process the complexities. Contrary to the criticized view of the privacy paradox (see Section \ref{narrow}), users have little choice than to accept obscure terms and conditions in exchange for online services either because the risks are not well understood or they are overburdened with information~\cite{solove2021}.

\paragraph{Lying} Requiring the possession of credentials as an obligatory point of passage for the provision of online services might lead to a situation where those who would not posses them or want to remain private, are incentivized to lie. Ramokapane~\etal{} outline the use of lying to access services online~\cite{ramokopane2021}. Page~\etal{} analysed survey data from 1532 participants and report the individuals who tend to lie  more ``have increased boundary preservation concerns as well as increased privacy concerns''~\cite{pagelying}. Lying to protect sensitive information and to avoid discrimination has been reported by Van Kleek~\etal{} in their investigation of reasons behind lying behavior online~\cite{vankleeklying2015}. One can debate around the reason individuals resort to lying~\cite{sannonlying2018}, but our concern is that many users perceive that the only effective means for them to preserve their privacy and/or dignity online is to lie rather than proper regulatory or technical protections.

%The tendency to lie leads to, for example, moral hazards in cyber-insurance~\cite{vakilinia2019}. To counter the effects of lying, service providers in turn resort to intrusion to ascertain the eligibility for access to certain goods or services. %It might be contextual to mention here that
%In such cases, the incentives are not only poor for the individuals but also for the entities that store the data. Healthcare providers too demonstrated reluctance to adopt electronic medical records (EMR); the adoption of which could have saved lives, particularly infant mortality. The complexities of the regulations acted as a barrier in case of the health care providers~\cite{miller2012}.%is this related to lying?

\paragraph{Reticence and fear} 
In the last decade, there has been increasing public awareness of the ubiquity of surveillance enabled by the huge amounts of data held by commercial actors (notably social media platforms) and the use of biometrics and face recognition technologies by governments. Numerous studies have evidenced the diverse and \emph{chilling effects} manifesting in practices of self-censorship, self-restraint or change of behaviours online such as limiting sharing of pictures or other information on social media ~\cite{manokha_surveillance_2018,forte2017}. 
Humbert~\etal{} conducted a survey on the interdependent privacy risks of individuals by the activities of their friends. Their findings highlight the impact of irresponsible online behavior of individuals on their friends or family who do or do not directly use technology~\cite{humbert2020}. 
These reactions limit the function of individuals because they might lose out on the benefits of participating in the digital economy or are afraid of expressing themselves freely. 
%\hl{There are instances where individuals with different levels of expertise failed to appropriate their \emph{functioning} of a private and safe life online because the system did not take into account their situations.} %not relevant to fear 
 
\begin{finding}
  \noindent
 An assessment of the abilities of individuals with diverse social and educational backgrounds to process complex technologies, legal documents and/or respond to ubiquitous connected devices, is missing in the evolution of PETs. The extant environment makes it difficult for many to achieve the \emph{functioning} of a private life.
\end{finding}
 
\subsection{Misplaced expectations about users}
The idea of provisioning of \ac{PETs} is often seen as a special benefaction for individuals, about whom several assumptions are made. Prevalent thinking among developers is toward \emph{fixing} the user. This is true in the way application developers perceive their users and how API providers perceive application developers (intermediate users). For both instances, the diktat is that users should behave in a particular manner else they are threats to the system. A mundane example is the futility of the expectation that users' would heed to SSL certificate warnings even though the over-use of warning messages has been criticised in that they can be counterproductive, thus defeating the purpose they were meant to serve~\cite{sassescaring2015}. 

\paragraph{The impact on users and non-users}The field of science and technology studies, and notably feminist scholarship, has been long concerned with how users are configured based on detached and self-referential models by developers (who commonly belong to privileged groups)~\cite{oudshoorn_configuring_2004, doorn_theorizing_2008}. Systems are typically built on tendentious assumptions driven by the point of view of said developers on what is \emph{right} for the user. This has led to inadequate generalizations manifesting in problems of misalignment (e.g. gendered technology) or exclusion on the basis of race, age, dis/abilities or other traits~\cite{oudshoorn_configuring_2004, doorn_theorizing_2008}. In the context of PETs, a prevalent assumption has been that users should be responsible for their own privacy which could be enhanced by means of anonymity, encryption and secure channels of communication~\cite{gurses_pets_2010}. Users are thus imagined as possessing the right set of knowledge, skills and resources to find and make use of PETs. However, there might be a disconnect between developers' \emph{idealized} assumptions about users and the very specific needs and identities of people at the other end of systems. This is particularly sensitive for those in high-risk, marginalized or vulnerable situations such as whistle blowers, victims of domestic violence, protesters or refugees~\cite{lauspeakers,brown2015social}. Another issue with predefined ideas of use is that developers may be biased towards those expected to interface with technology while being blind to a vast number of non-users (such as the elderly or disconnected), who may not directly interact with online systems but whose data may well be collected by various information systems~\cite{perera2015big}. Indeed, the frame of 'threats' is primarily concerned with the realm of the web or browser; it neglects the risk of being surveilled by other means such as sensors and IoT devices when such exposure leads to profiling, identification, discrimination and other dangers~\cite{solove_privacy, dominguez_hernandez_being_2022}. With the advent of big data, several ethical issues have come to the fore around the use of peoples' digital traces and statistical prediction for the automation of decisions related to access welfare, employment and public services, credit scoring~\cite{barocas_big_2016, ruckenstein_datafication_2017}. 

An analogy can be made between (threat) modelling thinking and the contractarian model of jurisprudence. Immanuel Kant and Rousseau proposed norms and legal institutions that would effectively compel citizens either to conform, or be outlaws. This approach is geared towards ensuring the survivability of the institutions rather than the wellbeing of the citizens they are meant to serve~\cite{senjustice}. In a similar way, the evaluation of systems based on norms operationalization and reductive models about complex human behaviour is usually aimed at fine tuning features that ensure the intended use of systems rather than on what real opportunities individuals have to use these systems~\cite{rikkebjergcivicdigital}. For privacy systems design, privacy policy is decided first, and the mechanisms to implement said policy are decided later. Anyone whose behavior deviates from the specifications of the systems designer is blamed ~\cite{mcdonald2020politics}. The protocol designer rarely provides reasons for their expectation of a particular allowed behavior. The problem with these assumptions is that they make systems rigid and unpleasant to use~\cite{pdc10}. \ac{HCI} research over the last two decades has argued strongly against \emph{blaming} and \emph{fixing} the individual towards being sensitive to their realities~\cite{adamss99,sassescaring2015}. The pandemic times have required people with various levels of backgrounds and deprivations to participate in online activities, for example, students from the poorest parts of the world as well as the elderly who might not be technically conversant\footnote{https://www.thelancet.com/journals/landig/article/PIIS2589-7500(20)30169-2/fulltext}. There is a clear need for developers of \ac{PETs} to study and build for vulnerable groups %upon users
who are less privileged, less abled or are in risky situations and who may be inadvertently rendered invisible during design. The extant pre-dominant approaches in human centered computing of interviews and focus groups are limited in their ability to capture the lived experiences of individuals~\cite{wang2018}. 

\paragraph{The impact on developers} This \emph{contractarian} attitude is also reflected in how security API developers relate to their primary users (developers). Applications developed with third party APIs and responsible for the privacy protection of their users often fail to do so. Hedin~\etal{} studied the flow of information through libraries provided by browser APIs and found that some sites ensured data does not leave the browser, or they only share it with the originating server. Meanwhile, others were freely propagating it to third parties~\cite{hedin2014}. Acar~\etal{} argue for a better understanding of the motivations and priorities of developers rather than blaming them for not being mindful of security. They stress the need for developer-centered studies to understand the challenges that developers face when using these APIs, and the resources available to improve the usability of these APIs~\cite{acar2016}. The APIs used by developers are not easy to use and to add to the challenges, the documentation to use them safely is not readily available or comprehensible. They sometimes interfere with the functionalities of the applications. 
\begin{finding}
  \noindent
\ac{PETs} have an expectation of a specified behavior, as well as adequate expertise from individuals who are supposed to benefit from them. However, individuals are \emph{active} agents, acting and doing things on their own. The gap lies in accommodating individuals who might not behave in a particular way as specified by the \ac{PETs} designer.     
\end{finding}

\subsection{Narrowly viewing individual attitudes towards privacy}\label{narrow}
A manifest shortcoming of failing to recognize humans in different contexts and cultures is the problematic academic view that individuals do not value their privacy based on their seemingly contradictory online behaviour. %For instance, based on a review of literature on privacy behaviour of individuals, Barth~\etal{} outline rational and irrational factors, lack of information and evaluation of risks that go into the privacy behavior of individuals~\cite{barth2017}. Acquisiti~\etal{} review the growing recognition of behavioral aspects of privacy decision making. 
The depiction of \emph{individuals} as rational beings albeit \emph{selfish} giving information in exchange for `insignificant' goods and services, led to the characterization of the privacy paradox~\cite{acquisiti2004,acquisiti2016}. This view however has attracted criticism over limitations in the rationale behind this concept where users are viewed as acting irrationally or not in accordance with their stated preferences~\cite{solove2021}. Having to sacrifice one's privacy in exchange for access to basic digital services or perceived benefits in the digital economy, might in fact speak more to the existence of coercive data collection systems than users acting incoherently. 
%The emergence of nudges to different types of users based on their preferences reflect a paternalistic tendency among the interventions~\cite{acquisitinudges2017}.
While the surface manifestations are studied in the context of HCI, as well as in the economics of privacy literature, we scrutinise if these manifestations are rational. Social choice theory is rich with research assuming rationality as one of many outcomes. When a wedding cake is cut some would want the icing while some the cake; however in most such cases individuals would seldom pick up the largest slice of the cake. This behaviour is inconsistent with the usual formulation with rationality as maximisation of \emph{selfish} interests. However, Sen describes it as 'menu-dependent behavior' given an individual's presumptions about how others will behave. Such behaviors broaden the scope of well being to include social traditions, imitation, as well as behavior driven by morality, sympathy and cohesion among others~\cite{senrational1994}. 
%these bit above is hard to read %I could not find a way to simplify this. I meant here with the example and terms used in social choice theory that individuals do act out of reasons like morality cohesion and those make them happy which is also a well being for them. Sharing the cake is well being more than eating it all by himself. However the constraint here is I have to use some terms and references they use in Social choice to establish this. 
When it comes to using online systems humans might have diverse yet perfectly rational reasons to trade-off their privacy (e.g., being overburdened with information or needing to access a service quickly, or just being kind and considerate towards others), which does not necessarily signify carelessness, naivety or indifference towards privacy.

\begin{finding}
  \noindent
  A distinct shortcoming of fixing users is to reduce the revealed preferences of individuals to simply reluctance or indifference towards privacy. The \emph{functioning} of a private life will need to assess how preferences are moderated by social dynamics. 
\end{finding}

\subsection{Power asymmetries and the creation of winners and losers} 
%People are at the center of the debates and mechanisms and technologies to support privacy. 
\ac{PETs} can potentially rearrange power between individuals %average end users
and large corporations/nation states who collect, store, process and benefit from information about their customers/citizens. %end users.
Yet, the formulation of a uniform set of requirements of privacy as universally beneficial for everyone across contexts is canonical and blind to structural inequalities and asymmetries~\cite{jacobchicago}. Such presumed uniformity across contexts fails to adequately account for diversity of circumstance and political reality. 
%For example Beyleveld observes that true anonymization in the context of medical research, can violate privacy rather than protecting it. In turn, the legal provision of the \emph{right to know} as part of privacy legislation can be compromised by anonymization techniques~\cite{chadwick2011}.
The multidisciplinary field of surveillance studies has extensively debated how commercial and political interests around surveillance  engender  multiple  ethical  tensions  and  creates winners  and  losers~\cite{andrejevic_big_2014,lyon_surveillance_2014}. While disclosing certain information in certain contexts may be deemed fairly unproblematic (e.g., to access students discounts), in more complex cases like criminal records, the degree of disclosure may directly impact equal opportunities for ethnic minorities~\cite{jacobchicago}. The situation is equally complex in the context of medical research. The absence of a transparent, verifiable data protection regime could directly affect legitimate and positive uses of data in medical research~\cite{nuffield}. On the other hand the perpetual nature of web ensures that misdeeds of individuals are permanently stored leading to discrimination based on past behaviour~\cite{brandimarte2015}. However, enabling individuals to delete their unpleasant past might be in conflict with economic and political interests such as national security, immigration and mobility policies, fraud prevention or policing. In practice, it is not easy for people to delete information they do not want to have in the public domain about themselves~\cite{ramokopane2017,MurilloKSL18}. Indeed, individuals as organised groups are politically weak to sustain the political pressure required for effective regulatory control and regime\footnote{The Moral Character of Cryptographic Work, Phillip Rogaway 2015 IACR Distinguished Lecture}. Politics remains extremely relevant in the effectiveness of regulatory controls and the benefits of a compromised control regime accruing to whom are pertinent questions in the realm of provisioning of \ac{PETs} as a social good.  
\begin{finding}
  \noindent
    The provisioning of technology like any public good is political and affects social groups. It is important to assess and understand if the consequences of provisioning specific \ac{PETs} at varying degrees spawns new kinds of disadvantages for some and/or advantages for others. This would be a reflection of the presence of deliberate influences, if any, on overall welfare interests and freedom in particular political contexts. 
    \ac{PETs} require appreciation of the individuals in their social, economic and political context; the ongoing tensions; and an evolved understanding of the winners and losers, they might end up creating.
\end{finding}

\section{A Brief Overview of the Capability Approach}\label{formal}
Sen proposed the \emph{Capability Approach} as a framework of thought and a formula to make interpersonal comparisons of welfare. The framework can be used to analyse well-being and poverty, liberty and freedom, development, gender bias and inequalities, justice and social ethics. Central to the \emph{capability approach} is an active individual with its \emph{beings} and \emph{doings}~\cite{sencapability}. The important primitives of the approach can be summarised as:
\begin{itemize}
    \item Functionings - Functionings are the \emph{beings and doings} of a person. For example, living a private life is functioning.
    \item Capabilities - This alludes to the idea of opportunity or advantage that an individual has, to achieve from the alternative set of functionings. It is a set of vectors of functionings.
\end{itemize}

There are two important constituents of the \emph{capability approach}. One is a list of \emph{basic capability}. Sen defines \emph{basic capability} as the ability to \emph{satisfy certain crucially important functionings up to certain minimally adequate levels}~\cite{seninequality}. An example of a \emph{basic capability} in the context of social welfare in particular geographies is avoiding premature death. The list is evolved through debate and participation, depending on the context, an example being the basic capabilities for gender inequality assessment in~\cite{robeyns2003}. The other important ingredient is the evaluation of opportunities individuals have to achieve those \emph{basic capabilities}. The reason being the mere possession of a good or service will not enable the functioning. What is needed is to have the skill, intelligence, physical ability, social and political environment -- capabilities to achieve a particular functioning. 

As way of illustration, let us consider that the ability to anonymously communicate over the Internet as a \emph{basic capability}. If we have two individuals who both lack the functioning of anonymous communication. Let us now consider if one of them is living under an oppressed regime and the other in a liberal society -- then from a political environment perspective, they would have different capabilities to achieve the functioning. So designing and provisioning of \ac{PETs} would need to be sensitive to their individual political realities. That said, other relevant factors beyond political environment that would influence the functioning would also need to be considered. For example, access to the Internet, health, education, ability, and so forth. In sum, the \emph{capability approach} can be effective to account for human diversities as it goes beyond the body and mind of the user to consider social and political conditions.  

In terms of formalizations, Sen~\cite{senformalization1985} and Robeyns~\cite{robeyns2001} presented the \emph{capability approach} as:
%\vspace{5pt}
\begin{mdframed}[backgroundcolor=gray!10,linewidth=0.1pt]

If \(x_i\) be the vector of commodities possessed by person \(i\) and \(c(x_i)\) converts the commodities into corresponding characteristics. The function \(f_i(c(x_i))\) converts the characteristics into functionings \(b_i\) s.t

\(b_i = f_i(c(x_i))\).

The function \(f\) is \(i\) specific because it depends on individual conversion factors, and each individual will choose a \(f_i\) of the set \(F_i\). Wang mentions that individuals with disability, vulnerability and the realities of socio-economically disadvantaged groups rarely find their interests and choices reflected in security and privacy mechanisms~\cite{wang2018}. This function \(f_i\) is the determinant in terms of physical and other abilities that Wang suggested to include for inclusive privacy analysis. A person with good health, nutrition and education will have a \(f_i \in F_i\) different from someone who does not.  

Robeyns extends the original formulation to account for social and environmental factors (e.g., policies, social norms, infrastructure) to be denoted as \(z_i\). Then the functioning

\(b_i = f_i(c(x_i,z_i))\).

For a given commodity vector \(x_i\), \(P_i(x_i)\) is the set of functionings feasible for a person \(i\) where \(f_i(.)\in F_i\).  

For any \(x_i \in X_i\) where \(X_i\) is the set of entitlements (commodities) the capability (or feasible functionings) \(Q_i\) is determined as \(Q_i(X_i) = b_i|b_i = f_i(c(x_i,z_i))\) where \(f_i (.) \in F_i\)  and \(x_i(.) \in X_i\).

\end{mdframed}
\vspace{5pt}

It is worth noting that the strength of the \emph{capability approach} lies in its attentiveness to context, diversity and choice -- which might not be adequately reflected, nor should it be lost in  formalizations of  \emph{capability approach}~\cite{robeyns2001,senformalization1985}
%Successful application of the \emph{capability approach} depends on its sensitivity to the context in which it is applied -- so a formalization, while helpful, should not be seen as equal to the \emph{capability approach}. 

\begin{significance}
We argue for an evaluation that will inform whether everyone is in a position to effectively benefit from the resources (i.e., \ac{PETs}), irrespective of their deprivations. The \emph{capability approach} offers an opportunity for designers and developers of PETs to build on a critical assessment and understanding of individual realities.
\end{significance}

\section{Towards a Situated view of PETs: An agenda for research and innovation}\label{agenda}
The preceding sections highlighted the need for a sufficient assessment of the real opportunities diverse individuals have to achieve the functioning of a private life. The \emph{capability approach} explicitly departs from welfare evaluations based on the availability of resources and/or policies. %and is based on ethical individualism. 
This emphasizes an assessment of individual abilities to achieve the \emph{functioning} in a manner they have a \emph{reason to value}. The advantage of this granularity is that diversity will not be subsumed under broad categorizations so as to preserve the interdependence among social groups~\cite{robeyns2003}. For example individuals who do not necessarily fit into `norms'~\cite{mcdonald2020privacy} would not be subsumed within the majority groups. In this section, we propose a research agenda aimed at making those assessments in a rigorous manner. 

%\label{Attending to capabilities}
\begin{figure}[!htp]
\centering
  \includegraphics[scale=0.5]{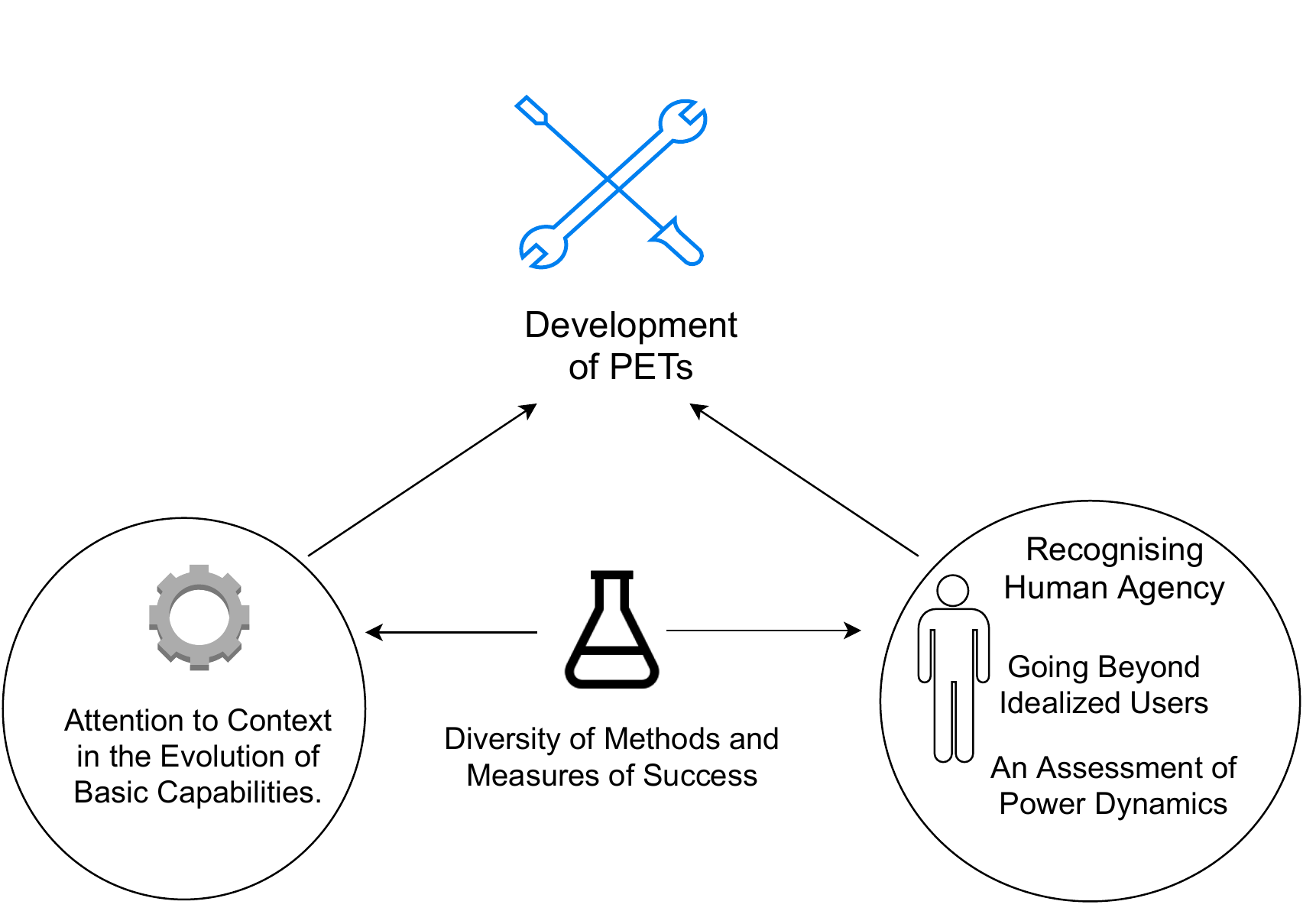}
\caption{The research themes with respect to the broad research agenda}
\label{high-level-design}
\end{figure}

Figure~\ref{high-level-design} gives an overview of areas of research that should be considered to embed the capability approach as the foundation of \ac{PETs}. The first area of research focuses on the evolution of the \emph{basic capabilities} that everyone should have. The second area of research aims to understand the individual the \ac{PETs} intend to protect. However, to fully achieve both elements, there is a need for novel methods and measures of success. Research should identify new ways of recognizing \emph{basic capabilities} and the metrics to qualify \ac{PETs} as successful and fit for purpose. We discuss these areas in detail below.

\subsection{Attention to Context in the Evolution of~\emph{Basic Capabilities}}
In the context of \ac{PETs} a basic capability can be the \emph{freedom} to perform basic actions online. We should note that \emph{basic capabilities} are not a definite list but should be understood with attention to context. The list can differ across populations with similar parameters of health, education, needs in different socio-cultural contexts. There are distinct groups ranging from migrants, to those living under oppressive regimes as well as citizens living in more liberal societies~\cite{guberek2018,gurses_pets_2010}. These political diversities, when juxtaposed with gender, race, education and other factors, can lead to a contextual granularity to a reasonable extent~\cite{schlesinger2017intersectional}. The list of \emph{basic capabilities} would determine the basic minimum protection mechanisms to which everyone should have access. 

For exposition, we refer to Solove's taxonomy; this is to give a shape to what we propose as \emph{basic capabilities}. Solove proposes four categories of harmful activities, namely (1) information collection, (2) information processing, (3) information dissemination, and (4) invasion~\cite{solove_privacy}. In light of this taxonomy and the harms discussed within each of these four activities, the list of \emph{basic capabilities} would need to be formulated in relation to how harms will provoke interventions such as (\ac{PETs}) to \emph{satisfy certain crucially important functionings up to certain minimally adequate levels}. A \emph{basic capability} example would be the ability to access state welfare benefits without being subjected to unauthorized disclosure--for instance, surveillance when inferences (whether accurate or not) are made about individuals and groups to inform decisions that affect their lives~\cite{forte2017, suchman_algorithmic_2020}. Furthermore, for disabled citizens, the \emph{functioning} will be further granulated based on their interface with technology. 

The list of harms can be informed by frameworks that elicit the threats and risks associated with the distinct scenarios and interactions that citizen groups have with systems~\cite{linden2020}. 
A pertinent issue is if developers would do this -- this is where positioning of the \emph{capability approach} is important. We discuss the alternatives of situating it in the policy layer or with developers/designers further in Section \ref{conclusion}.  %I think this section is lacking examples on how things play out differently in different contexts to illustrate why research should be situated.  
Yet, because evaluations need to be situated, we pose the formulation of \emph{basic capabilities} as an open question to be considered by researchers and practitioners. Prescribing a list risks presuming an antecedent uniformity and that there is a \emph{right} method. Furthermore, the process in which a list of capabilities is evolved is very critical from the perspective of the \emph{capability approach}. Nonetheless, it is pertinent to draw attention to the debates among scholars working in social justice and welfare, for a definite list against evolving a context-dependent list~\cite{martha1988-NUSNFA-2,nussbaum_2000}.

%this sounds a bit contradictory, if we bring attention to context, a 'definite' list seems to counterpose that idea. I draw this to counter the solove's list that is why solove's list came in the next sentence. We are not bringing context merely pointing out to the debate for people who would want to implement this or know more. 
\begin{research}
 Future research can deliberate whether existing propositions like Solove's taxonomy or the LINDDUN framework~\cite{linddun2018} are adequate or if a nuanced contextual list should evolve through broader participation. A starting point would be to explore the extent to which the existing recommendations are in synergy with the political and economic maturity of various geographies as well as their cultural and social histories~\cite{sencapability}. 
\end{research}

%Crucial to a more plural conceptualization of PETs is the need for more nuanced categories and vocabularies which are sensitive to vulnerable groups or any actor implicated in systems of data collection which, by their situations, may not fit nicely under the conventional definitions of \emph{users}.

\subsection{Going beyond idealized users}
However prevalent in the spheres of technology development, the term `user' is reductive as it may unhelpfully gloss over diverse social realities and socio-technical relations. We need more nuanced categories to represent marginalized, vulnerable groups who do not fit into the conventional definitions of a user. A consideration of being inclusive of diverse observed abilities, needs and circumstances would require accepting them as legitimate focal variables as opposed to naive assumptions of universality. For example, individuals with different degrees of education should not be deceived by complex legal agreements. %This one sentence is showing our concern for individuals with diverse skills. So I am inclined to keep as that is a point we make in information overload and the rest of this agenda is about surveillance etc. 
The plurality of focal variables means that there could be multiple conceptualizations of \ac{PETs} and other privacy protections, in terms of distribution, participation, abilities, and changing circumstances; at both the individual and collective level. 

There is a need to expand the scope of action of privacy protection mechanisms to attend to different socio-technical arrangements and human relations with technology. Designers ought to ask who will benefit from the enhanced privacy protections of a particular design, and who will not. Such assessments concern ethical questions of inclusivity, dignity, and justice~\cite{jacobchicago}, which call for a focus on the disadvantaged and their often invisible realities. Recognizing the heterogeneity of socio-technical relations will enable different groups to enjoy social good in a manner they can, neutralizing the limitations (if any) of their opportunities to do so to a reasonable extent. For example, a recent study on online safety settings for couples with memory concerns highlights their need for flexibility without fundamentally altering their relationship~\cite{mcdonaldmemoryconcern}. 

The development of PETs should turn to approaches in sociology such as intersectionality~\cite{schlesinger2017intersectional} and the well established practice of reflexivity within qualitative research whereby personal biases, assumptions, motivations and institutional commitments can be made visible, if not recognized as potential limitations of design~\cite{pihkala_reflexive}. In her work on `locating accountabilities in technology production', Suchman has argued for a shift from flawed `view from nowhere' ideals toward locating design --always from `somewhere'--in synergy with different ways of being and partaking in technology design~\cite{suchman_located_2002}. We highlight a few productive efforts which have started to look at ways to attend to the specific security and privacy needs of high-risk or vulnerable individuals such as whistle blowers, protesters, and refugees~\cite{ermoshina_can_2017, simko_computer_2018}. % %how could this be done is not clear. I mention their framework in the methods. Maybe this sentence belongs in the previous subsection----- Taken
Caring for the privacy of both users and non-users will help recontextualize the role of PETs beyond the interface, for example, in response to ubiquitous and pervasive technologies or the various layers of technical systems involving data collection, transport, and processing~\cite{dominguez_hernandez_being_2022}. 
The \emph{capability approach} seeks to enable plural conceptualizations of citizens through assessing the diverse abilities of individuals to operate systems, process the risks based on their knowledge and sensory abilities, and interactions (or not) with technology, leading to wider exercising of the \emph{functioning} of a private life.

Our proposition for a nuanced understanding and delineation of \emph{users} requires a broad understanding of the remit of designers/developers. The conceptual framework to operationalize the \emph{capability approach} needs to be contextual with the explicit goal of avoiding generalizations that subsume human diversity. Moreover, discussion about the remits and responsibilities require understanding the distinction between a public good and provisioning of the same public good.  Consider, for example, the case of a technology to anonymously browse the web. If agreed to be beneficial to all browser users (a desired \emph{basic capability}), then how the tool is provisioned will determine if everyone is being able to use it in a manner they can and they value -- design of the access to the relevant \emph{basic capability}. Such a tool should be decoupled from market considerations and it would be in the remit of internet browser designers and developers to provision it by default. %A study on the use of Tor and JonDonym highlighted browsing compatibility as a key consideration for developers -- this means that privacy protections should be accessible to websites users regardless of the site they would want to visit~\cite{rannenberg2020}. Ideally, \ac{PETs} would not lead to any noticeable limitation on the quality of service being accessed~\cite{harborth2017integrating}. However there might be technical trade-offs in attempting to serve specific needs. While compatibility (or interoperability) with websites is at the back end and can be uniform across all users --the requirements at the user-facing side would differ. 
Because the \emph{capability approach} does not presume uniformity, designers/developers will need to attend to specific scenarios and needs (e.g., protecting activists or journalists from state surveillance) where the abilities, needs and expectations will differ~\cite{herrmann2015anonymity}. 
In fine, the mandate for the designer would be to adhere with the agreed \emph{basic capability}, while the remit would be contextual on their target user group. Nevertheless, there will be overlaps between user groups and they are not deviations from the conceptual framework of the \emph{capability approach}~\cite{robeyns2003}. A pertinent question is whether contextuality harms interoperability -- our view is that the latter is a back-end systems property~\cite{framework2004european} and can co-exist with contextuality. Furthermore, the \emph{capability approach} is not restrictive of exclusive services beyond \emph{basic capabilities} keeping specific sections of society in mind.

\begin{research}
Research should focus on a nuanced and systematic understanding of the diverse realities of the individuals that \ac{PETs} intend to protect and the challenges faced by developers to acquire and act upon such understandings. More apt terminologies, beyond monolithic categories such as ‘users’, are needed to allude to the intended beneficiaries of PETs in all their diversity. On the other hand we recommend that developers engage in self-audit, reflective practices aimed at making assumptions explicit. 
\end{research}

\subsection{Recognizing human agency}
A key consideration of the \emph{capability approach} is that individuals should be able to achieve the \emph{functioning} of a private life in a manner they have a \emph{reason to value}. This puts \emph{agency} at the heart of \emph{functioning}. Individuals reveal information to remote entities and trust them to prevent identification, exposure, and other threats to misuse of the information~\cite{sokhateabuse}. Fears of misuse of sensitive information can lead to reticence or lying~\cite{ramokopane2021}, where individuals act driven by morality, compassion, and less self-centered views of rationality. We argue for allowing self-selection by individuals as a possible alternative to a supply side decision of what is good for them. For example, individuals can choose to share information for medical research provided they are explicitly beneficial and governed by morally appropriate authorities~\cite{nuffield}. In other applications, an individual may willingly subscribe to receive advertisements for certain products without that being an indication that the individual does not value their privacy. A possible self-selection approach and a potential remedy against individuals having to lie (e.g., giving false email addresses) could be to allow them to have an anonymous account with minimum personal data or which
cannot be traced back to them~\footnote{For this exposition, we refer to a recent initiatives by \emph{DuckDuckgo} and \emph{Apple} to allow users to hide their email addresses by redirecting emails based on preferences, and only those desired by the users will be delivered to themhttps://www.theverge.com/2021/7/20/22576352/duckduckgo-email-protection-privacy-trackers-apple-alternative}; with the caution that this type of solution will still require users to trust an intermediary (i.e., Apple or DuckDuckgo).
%Entities other than \emph{DuckDuckGo} will not be able to trace an email address back to an individual~\footnote{https://www.theverge.com/2021/7/20/22576352/duckduckgo-email-protection-privacy-trackers-apple-alternative}.

While the GDPR Act 6(1)\footnote{https://eur-lex.europa.eu/eli/reg/2016/679/o} states that \emph{the data subject has given consent to the processing of his or her personal data for one or more specific purposes}, instances of violations~\footnote{https://ico.org.uk/media/action-weve-taken/enforcement-notices/2620027/emailmovers-limited-en.pdf} brings to the fore the dangers of sharing more information than is required. We are not asserting that making such decisions are within the cognitive load and cognitive capacity of all individuals~\cite{solove2021,colnagocognitive2020}; we are recommending a more nuanced understanding and representation of the contexts and proportionality thereof. Merchants who violate regulations keep on collecting more data than is required, hiding behind complex consent controls or the opacity that separates end-users from merchants. Such understanding can potentially influence the implementation of the law in both letter and spirit and eliminate excessive data collection right at the point where individuals actively or passively interface with online systems. 

Van Der Linden~\etal{} explores software developers' attitudes towards the collection of data from their users. They find that developer's attitudes are not guided by the established principles of being `adequate, relevant' and `limited' to the purpose for which the data is collected~\cite{linden2019}. This conclusion is being arrived at by the authors while evaluating against specific regulations which might or might not be in sync with what citizens would prefer. 
\begin{research}
We recommend further research to empirically understand individual \emph{choices} to give an evaluative understanding of the interactions citizens have a \emph{reason to value}. Our recommendation is for a rigorous understanding of people's choices, intentions, values, and motivations, irrespective of what developers/regulators think are good for citizens. This evaluation can feed into negotiating the proportionality of information disclosure
particular to contexts and inform regulations/systems.
\end{research}

\subsection{A reflexive assessment of power dynamics}
While a disciplined assessment of the deprivations, valued interactions of active individuals should form the basis of \ac{PETs} is an insufficient requirement. Any technical development is not a self-contained exercise but is largely contingent on power structures and what the political and economic forces would be willing to concede. As feminist scholars have shown, the spheres of development have been largely dominated  by privileged groups who, despite good intentions, are unable to properly address the experiences of the oppressed, marginalized, disabled or vulnerable. Moreover, organizational affiliations and the exigencies of funding institutions, be they private or public, will constrain developers' scope of action.

There are applicable experiences from provisioning public goods in their disproportionate use and availability among the population--the ability to appropriate operates at many layers. In September 2019, the Court of Justice of the European Union (CJEU) ruled in two cases (C-136/17) and (C-507/17)~\cite{globocnik2020}. In the former, while the court made an implicit acknowledgement of the right to be forgotten, in the latter, the same court limited the territorial scope of the same right. Since CJEU nudged the lawmakers to consider expanding the territorial limits of GDPR, the way forward is driving public opinion for the lawmakers to take it up with their counterparts in other jurisdictions. Google is a profit-making enterprise making use of and profiting from the information they store about individuals. A pertinent question thus relates to the prudence of entrusting Google to decide which information is in the public interest and which is not. The rise of the data economy has put corporations under mounting regulatory scrutiny when it comes to accommodating public interest which is at odds with profits~\cite{zuboff_age_2019}. 

The other issue concerns the ability among various groups to use a public service when it is available. Several factors engender the widespread uptake of such services, however, a significant contributor to ability is the awareness among citizens of their rights and recourse to violations. The experience is not encouraging among vulnerable sections of society for access to justice in general~\cite{gill2021}, and when there is access, the battle is far too long and draining\footnote{The case of the UK post office miscarriage of justice is a good example. See https://www.postofficetrial.com}. The information asymmetry does not exist by itself but sometimes by bureaucratic design~\cite{hood1995}. A strength of the \emph{capability approach} is that along with the explicit consideration for human diversities; it actively factors in political realities as a critical conversion factor for individuals to lead the life they value.  

The design of privacy protections should not only recognize the heterogeneity of individual abilities and needs, but interrogate who is (and should be) in a position to devise and recommend said privacy protections without conflict of interest, and which regulatory interventions and political supports are needed to further the technical goals of PETs. In the data economy, the provisioning of privacy protections should be free from the influence of actors who profit, directly or indirectly, from more data collection. Not only that, capability-informed privacy protections cannot, by definition, be subject to payment or tiers of exclusion that would lead to a situation of privacy haves and have-nots. These fundamental tensions demand self-critical reflection about the limitations of designers and developers, the need for more inclusivity in the spheres of research and innovation, and nuanced recognition of the influence of deliberate political and economic forces that can limit what can be achieved in practice.
\begin{research}
The political economy of privacy protection foregrounds that technological responses should not be viewed as a panacea where too much energy is put into driving adoption and carrying out continued usability improvements. The design of privacy protections should not only recognize the heterogeneity of individual abilities and needs, but interrogate who is (and should be) in a position to devise and recommend said privacy protections, and which regulatory interventions and political supports are needed to further the technical goals of PETs. This calls for a reflexive exercise of the limitations of developers, calls for more inclusivity in the spheres of technical development, and nuanced recognition of the influence of deliberate political and economic forces that can limit what can be achieved in practice. 
\end{research}

%Still needs some concrete recommendation
%ways to enable political support for, and feasibility of, aiming \ac{PETs} specifically at enabling the individual to achieve the \emph{functioning} of a private life.

\subsection{Diversity of methods and measures of success}
A shift from the supply side view of what citizens need to more downstream, plural, conceptualizations of individuals will bring in cogency and make their participation in online activities enjoyable and valuable. The method one adopts to realize the research agenda is crucial to the success of embedding the \emph{capability approach} as a foundation of \ac{PETs}.  

\paragraph{How to prepare the list of basic capabilities?}
The process by which the list of \emph{basic capabilities} and interpersonal comparisons will evolve is crucial for the \emph{capability approach}. Such a list is significant for policy evaluations or measurements related to privacy (or lack thereof). The legitimacy of the list is critical in effecting \ac{PETs} as a means of social justice and democracy. Sen explicitly recommends debate and democratic participation to evolve the list. Selection will be an inescapable part of this process; which would mean catering to the needs of particularly vulnerable groups, in terms of ability and/or education and environment. Contemporary political philosophers have been engaged with the issues concerning selection in other contexts; we refer to the work of Robeyns for exposition~\cite{robeyns2003}; however, we are not rigid about a particular set of criteria. We briefly outline the criterion Robeyns used to evolve a list of \emph{basic capabilities}.  \emph{The criterion of explicit formulation} and \emph{the criterion of methodological justification} requires that the selected \emph{basic capabilities} should be defensible on both these counts. The list is required to be sensitive to the \emph{context} of the target group. \emph{The criterion of generality} specifies that the list should be evolved in two stages. First, a general list and second, a \emph{fine grained} list will be drafted enumerating all the \emph{basic capabilities} a citizen should have. This list will be refined based on local conditions based on data and empirical research. It is important that selected \emph{basic capabilities} might only have negligible overlaps with others to satisfy \emph{the criterion of non-reducability}~\cite{robeyns2003}. 

\paragraph{Methods to include individuals with diverse abilities and situations}
Though \emph{human-centered design} (HCD) has dominated conversations as an approach to promote increased adoption and use of systems, this has been often reduced to user studies and consultation~\cite{buchanan2001}. Moreover, there is a limit to what developers can learn from their users, given several social, material, and political constraints~\cite{stewart_wrong_2005}. This owes, among other factors, to varying degrees of ignorance and technical literacy, issues of accessibility (cognition, location, vulnerability, language, information overloads), and the presence of vast information asymmetries between users and highly opaque information systems. In re-imagining human-centricity, the \emph{capability approach} entails much more than the notion of \emph{utility} implicit in usability~\cite{oosterlaken2009} -- a preference ordering of satisfaction with regards to surface features. There is a moral obligation of \ac{PETs} to cater to those whose ``body and mind" do not fit the conventional construction of a \emph{user}. Observed diversities and realities are as crucial as those that are unobserved. We borrow the term \emph{unknown known} from~\cite{rashid2016} to emphasize the emergent and continuously evolving nature of individuals and the environment. Focus groups, interviews, and other participatory research methods have proven highly productive, yet they need to be cautiously implemented so as to avoid exploitation and burden~\cite{pierre_getting_2021} and allow for more generative spaces to understand issues of marginalization and evolving environmental realities. We advocate for more inclusivity in the spheres of design and alliances with methods from social sciences as a means to develop better interventions. For example, Albrecht~\etal{} conducted ethnographic research with 11 protesters from Hong Kong to understand the improvisations and unusual tactics protesters resorted to in order to avoid state surveillance~\cite{albrecht2021}. Schlesinger~\etal{} introduce the sociological framework of `intersectionality' in HCI to understand the complex identities and experiences with marginalization of individuals. While they acknowledge the progress made in unpacking questions of identity, they also point out the gaps in addressing multiple forms of exclusion and oppression based on gender, race or class~\cite{schlesinger2017intersectional}. Such conceptual and methodological frameworks can feed into the normative evaluations of the conversion factors of individuals to achieve the \emph{functioning} of a private life in similar situations.%While we give examples of ethnographic studies, frameworks to argue for fresh approaches of participatory research, the richness of human diversity can be the foundation to inform the methodologies to evolve a \emph{capability} centered design. 
\paragraph{How to measure success of adopting the capability approach?}
While we depart from the comparison of welfare based on possession of resources, the critical question is how do we propose to evaluate \ac{PETs} built using the \emph{capability approach}. %A potential indicator is \emph{functioning} keeping \emph{choice} at the very heart of it~\cite{robeyns2001}.
Conventionally, technologies have been mainly measured in terms of their adoption or acceptability which cannot always account for unexpected uses and reactions, or the effectiveness of the technology to live up to its promises. Future research can delve into the metrics and assessments which are not merely techno-centric but can more adequately reflect how citizens are able to exercise their \emph{functionings} and enjoy their right to privacy. \emph{Functionings} can be observed not only quantitatively but qualitatively, for example, if a journalist living under an oppressive regime can exercise their right to a private life without oppression. Measures of success should factor in the diversity of beneficiaries according to their situations and complex identities with respect to race, age, ethnicity, gender, sexuality, social and political realities, physical handicap, mental health, pregnancy, or have caring responsibilities. While factoring in diversity, adequate care should be taken to limit the discrimination among users and exclusion of non-users.  
\begin{moral}
We do not advocate an explicitly reductionist~\cite{andersonreductionism} approach in operationalizing the \emph{capability approach} by applying laws/results from one discipline to another. The inherent scale and complexity of human diversity and technology respectively would reveal new assumptions, needs and compromises. These new assumptions, needs and compromises are as fundamental to our discipline as their counterparts in other disciplines. Adoption of the \emph{capability approach} to address them for \ac{PETs} would require as much rigor as any other discipline. 
\end{moral}

\section{Case Illustration}\label{why} 
In Section \ref{forwhom} we make a case for an informationally adequate approach to serve as the foundation of \ac{PETs}, describing the \emph{capability approach} and proposing a research agenda in Sections \ref{formal} and \ref{agenda} respectively. Here we ground our framework by making a preliminary assessment of its implications on specific cases -- this we term as case implication justification\footnote{Sen coined the term case implication critique while referring to the shortcomings of utilitarian utility and Rawlsian equality in~\cite{seninequality}}.

The National Cyber Security Centre UK (NCSC) annual review, 2020 highlights that many cyber security attacks
can be prevented through simple steps. However a considerable proportion of the public are often found reluctant to take those steps\footnote{https://www.ncsc.gov.uk/news/annual-review-2020}. We explore the safe social media usage guidelines published by NCSC\footnote{https://www.ncsc.gov.uk/pdfs/guidance/social-media-how-to-use-it-safely.pdf}, particularly the guideline on digital footprints in the Section ``Understanding your digital footprint", as a case in point to explore:
\begin{itemize}
    \item Can individuals take those steps in a manner they can and they value?
    \item Where and how do we situate capability approach?
\end{itemize}
We restrict our discussion of digital footprints with respect to social media in this paper. 
\paragraph{Can individuals take those steps in a manner they can and they value?}
An important guideline suggested by NCSC states:
\begin{quote}
    ``Think about what you're posting, and who has access to it. Have you
    configured the privacy options so that it's only accessible to the people you
    want to see it?"
\end{quote}
McDonalds~\etal{} explored the privacy narratives set by large social media companies which have a deliberate influence on the privacy features and controls available to their users~\cite{mcdonaldsm2021}. Their findings report that Snapchat creates a false sense that user data is ephemeral through a misleading description of the self. The authors also identify the misleading assertions made by Facebook regarding data shared with friends without explicitly stating the same is shared with advertisers as well. Summarising, the authors report that the large companies are conservative with the truth, confusing users with hard to comprehend terms and create an illusion of control and power. A study by Marwick~\etal{} highlights the helplessness felt among socio-economically disadvantaged youth when it comes to privacy in the networked world; they are the most susceptible to privacy violations yet they are unable to comprehend the dangers to which they are exposed~\cite{marwick2017nobody}. In 2019, when deciding on whether a person has the capacity to decide on their internet and social media use, a judge in the United Kingdom observed~\footnote{https://www.bailii.org/ew/cases/EWCOP/2019/3.html}: 
\begin{quote}
    I do not envisage that the precise details or mechanisms of the privacy settings need to be understood, but P should be capable of understanding that they exist and be able to decide (with support) whether to apply them.
\end{quote}

The other pertinent issue is whether individuals can exercise those steps in a manner they value -- this entails the notion of agency~\cite{senrational1994}. The continuous advancement of norms by the large social media companies as cultural adaptations leads to an environment of exclusion for those who do not fall into these norms~\cite{mcdonald2020politics}. The equation of norms to cultural adaptations subsumes the nuances of human diversity like vulnerable personal identities, as well as groups like socio-economically disadvantaged, refugees and the persecuted in oppressive regimes. Majoritarian norms as cultural adaptations do not augur well for those in the minority in social media platforms -- for example, Facebook discounts bad behavior such as targeted bullying as a `natural outcome of social interactions~\cite{mcdonaldsm2021}'. Moving on from individuals to the environment, the cases of abuse of human rights and freedom of the press in various parts of the world are well documented\footnote{https://commonslibrary.parliament.uk/research-briefings/cdp-2020-0063/}. Citizens in some parts of the world, even when equipped with the resources (e.g., devices and the Internet), are not able to exercise their \emph{functioning} of private life and freedom of speech. Citizens live in an environment where they are profiled and watched without their knowledge or consent, even when they do not directly interact with technology. The prominent social media narrative of ``nothing to fear" if conformed to ``acceptable" behavior norms, leads to serious persecution of activists and political minority groups living in oppressed regimes\footnote{https://reutersinstitute.politics.ox.ac.uk/news/religious-riots-grow-india-critics-accuse-facebook-fanning-flames}.

\paragraph{Where and how do we situate the capability approach?}
This will be a brief theoretical exploration of the deployment of the \emph{capability approach} to enable social media privacy for individuals in a manner they can and they value. 
\begin{enumerate}
    \item \textbf{Basic capabilities} - The list of \emph{basic capabilities} forms an important pivot of the \emph{capability approach}.  A key requirement of the \emph{capability approach} is that such a list should evolve through public participation and democratic means. There are suggestions of using legal reasoning~\cite{woods2021blessed} to evolve cyber security controls -- the authors contend that ``Controls will be prioritised based on reasonableness or appropriateness
    rather than effectiveness". The distributive considerations of the \emph{capability approach} can serve as an important ingredient in the paradigm of ``law inheriting cyber-security" to consider matters of realised justice as opposed to only law~\cite{senjustice,pdc10}. 

    \item \textbf{Understanding users} -- We delineate \emph{going beyond the user} and \emph{recognizing human agency} as two key research agendas in the realization of the \emph{capability approach}. With respect to our case of social media protections against digital footprints, the \emph{capability approach} points to the information on social norms, habits, age, education and ability to make an evaluation of well being and deprivations. The individual, their freedom and choices are at the heart of this assessment. This information can be used to create personas~\cite{lewis2014says} that represent those very individuals for whom the privacy controls are being designed. Cyber security narratives can be created to explain protection mechanisms to the personas in a manner that they understand, will be able to operate and that they would enjoy. An effective example of narration in cyber security can be referred in~\cite{liveleystories}. 
    
    \item \textbf{A reflexive assessment of power dynamics} -- In our extant case, understanding human diversity is not enough due to the power asymmetry that exists between individuals and large social media companies. West examines the rise of data capitalism and the narratives that have always been built to propagate the appropriation of data, often at the cost of individual rights -- language played an important role here~\cite{west2019data}. This is further exacerbated when the less powerful are not conversant with the language with which power operates~\cite{bourdieu1991language}. We propose that the \emph{capability approach}, with its informational richness on \emph{basic capabilities} and \emph{human diversity}, be referred by the regulator to set the narrative so that individuals are not exploited through a false sense of security and benefit.      
\end{enumerate}

\section{Conclusion}\label{conclusion}
\begin{figure*}[!thp]
    \centering
    \includegraphics[width=2.1\columnwidth]{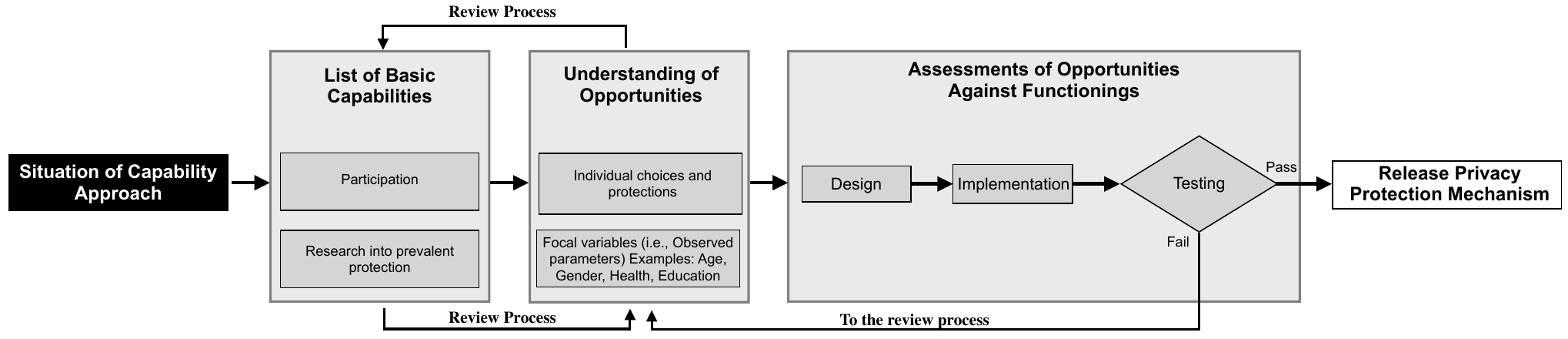}
    \caption{Stages of Implementation}
    \label{fig:steps}
\end{figure*}
In our view of using the \emph{capability approach} as a foundation for building \ac{PETs}, we recognize the moral obligation of any privacy protection mechanism to human agency and diversity.  This view attends to the need to cater for individuals in all their complexity, and advocates for empathy, accountability and transparency in the process of development. Much in line with Suchman's critique of objectivist `design from nowhere' claims~\cite{suchman_located_2002}, we advance that the design of privacy protections should be reflective of partial views and institutional limitations, while sensitive to the plural realities of users and non-users of technology, their diverse needs, locations, lived experiences, identities, preferences, abilities, and social and environmental conditions. This can be achieved by foregrounding developers' commitments and preconceptions and inviting all stakeholders not only to deliberate but to conceive better ways to protect privacy. It is important to acknowledge that in any human-centered approach, the ability to garner privacy requirements is encumbered by limits in the knowledge possessed by the selected cohort of users and their actual means to inform designers on what is needed~\cite{stewart_wrong_2005}. Overlooking these constraints could be detrimental if it creates a false sense of certainty and human centricity. As recent studies on surveillance have shown, privacy violations linked with algorithmic behavior prediction are highly complex and can occur completely unbeknownst to people even when \emph{good} legal privacy provisions and technical measures are in place~\cite{miller_total_2014}. Because of this, more attention has been given to ethical issues associated with the datafication of human activity and its use for statistical inference and prediction of future behaviour~\cite{muhlhoff_predictive_2020}. Opacity not only makes locating responsibilities difficult but unavoidably creates a situation of unevenly distributed costs and benefits. This underpins our emphasis on promoting reflexivity to surfacing power asymmetries and structural inequalities in technology design and development. Our proposal calls for a bottom-up view of citizens in their environment (particularly those in vulnerable situations), which aims to expand the repertoire of empirical methods to inform socio-technical interventions.

We end with a high level view of the steps to enable a bottom-up view of citizens in their realities as in Figure \ref{fig:steps}. 

\begin{enumerate}
    \item \textbf{Situation of capability approach} -- This we believe is an immediate step -- meaning whether it should be situated at the level of technology policy or at the level of implementation of technology. The former would mean the interventions at the level of policy leading to comprehensible guidelines for implementations.   
    
    \item \textbf{Method of evolution of list of basic capabilities} -- Once we ascertain the placement of the \emph{capability approach} the next step can be the method to evolve the list of \emph{basic capabilities} -- the minimum functionings every individual should have. While there are examples between a definitive list against a more contextual ones -- the method will output a list that is `appropriate' and `reasonable'.  
    \item \textbf{Understanding of Opportunities} -- One can refer to the work of United Nations Development Program's adoption of the \emph{capability approach} to food security for exposition on the granularity of data collection~\cite{burchi2012human}. This mixed method step includes:
    \subitem - determining the focal variable(s) -- would we assess individuals based only on their education or include more observed diversities. This would range from education, ability, gender, age to cultural and religious beliefs; 
    \subitem - agency -- ``the ability of people to help themselves and also to influence the world"~\cite{sen1992};
    \subitem - political and social environment -- an understanding of the individual in its circumstances.
    \item \textbf{Assessment of opportunities against functioning} -- Once we have a understanding of opportunities, this is used to evaluate them against the functioning of, for example, private life. Here, cultural beliefs would indicate their language and communication preferences, along with their privacy beliefs.    
\end{enumerate}
The steps we present here are for exposition and are not comprehensive -- drawn from the applications of the \emph{capability approach} in other domains. 

\begin{acks}
We thank the anonymous reviewers, Ola Michalec and our shepherd Tara Whalen whose comments helped improve the paper greatly. This work is supported by REPHRAIN: National Research centre on Privacy, Harm Reduction and Adversarial Influence online (EPSRC Grant: EP/V011189/1).
\end{acks}

\bibliographystyle{ACM-Reference-Format}
\bibliography{PEPP}

\end{document}